**Role played by port drains in a Maxwell fisheye lens**

**Shen Q. (沈祈茵)[1], Gu C. (辜朝洁)[1], Li M. (李明)[1], Zhang X. (张璇) [1], Xiong H.[1], Liu Y. (刘泱杰)[2,1,3,*], Jin L.[2], Wen M.[2], and Wei Z.[2]**

1. School of Micro-Electronics, Hubei University, Wuhan 430062, Hubei Province
2. School of Physics, Hubei University, Wuhan 430062, Hubei Province
3. Lanzhou Center for Theoretical Physics, Key Laboratory of Theoretical Physics of Gansu Province, Lanzhou University, Lanzhou 730000, Gansu Province

[*]Corresponding author: yangjie@hubu.edu.cn

(Coauthors' emails: 201931105020102@stu.hubu.edu.cn, Guchaojie2102@qq.com, hubuliming@qq.com, Zhangxuan00603@qq.com; xionghui@fudan.edu.cn, jinluling@hubu.edu.cn, wenmeng@hubu.edu.cn, weizhengrong@hubu.edu.cn )

**Abstract.** Maxwell fisheye lens was proposed to reach super-resolution with the addition of wave drain, and the interaction of multiple drains is theoretically predicted to improve subwavelength resolution further. In this paper we discuss the role played by port drains in optical absolute instruments, and verify by wave simulation that coupling nature for wave source and drain applies correctly in the picture of scanning imaging for absolute instrument. This work prospects for scanning near fields shaped from far-field wave propagation.

Condensed Abstract *(optional)* Maxwell fisheye lens was proposed to reach super-resolution with the addition of wave drains, and the interaction of multiple drains is theoretically predicted to improve subwavelength resolution even further. In this paper we discuss the role played by port drains in optical absolute instruments, and verify that coupling nature for wave source and drain applies in the picture of scanning imaging for absolute instrument.

# 1. Intro.

Self-imaging is the fundamental concept behind the optical absolute instrument in wave imaging [1] as compared to other techniques such as negative-index superlens [2] and fluorescence microscopy [3]. This self-focusing imaging behavior was born out as a surprising result of transformation optics [4-6]. Previous progress [7] found that the conformal mapping [8] transforms multiple points to one in two-dimensional space, thus peculiarly making an absolute instrument device enabling self-imaging similar to Ref. [9]. Therein a super-resolved image occurs only when one looks at it via the wave drain [10-12]. However, the wave drain issue suffers heavily from yet-elusive understanding of the wave focusing in optical absolute instruments such as Maxwell fisheye (MFE) lens [13], and previous theoretical insight points out that full numerics to resolve the super-resolution remains difficulty due to the inordinate scale difference across between the drain and the imaging device [14]. Nevertheless, a simulation work adopts the wave absorber as a possible prototype to realise the wave drain effectively [15, 16] and we move forward with a simple tunable wave port as either source or drain in the scanning process of absolute instrument, in the hope to pindown the universal role of wave drains in benefit of subwavelength imaging [17, 14], which issue then becomes the purpose of our paper.

In this paper, we demonstrate by wave simulation the role played by port drains to induce subwavelength imaging, which accords with a universal drain theory [17]. Once this issue is resolved, the enigma of perfect imaging with drains in absolute instruments shall be clarified further to benefit the design for super-resolution imaging devices.

## 2. Absolute instruments and its super-resolution imaging

In Sec.2, we shall briefly convert Mikaelian lens to the generalized MFE lens for the sake of clarity via the classical conformal mapping [7]. Then we investigate the role of wave drain/sink to contribute to subwavelength imaging.

We first demonstrate the MFE profile via conformal mapping a Mikaelian lens. Conformal mapping refers to mapping a two-dimensional region to another two-dimensional region by an analytic complex function [8]. It was known to provide isotropic transformation media which gives rise to non-reflective wave propagation due to its superior angle-preservation property. A recent paper [7] found that a kind of isotropic inhomogeneous medium can be converted to an absolute instrument by using conformal mapping, which is enabled by the multiple-to-one mapping of conformal functions, for example, Mikaelian lens [18, 7] as an example to obtain other new absolute instruments. We will focus on the drain role to implement subwavelength imaging both in MFE lens and other absolute instruments. The refractive index of z-space is related to that of w-space as $n_z = n_w |dw/dz|$ where a conformal function $w(z)$ connects both spaces in a faithful manner. With the refractive index profile of the Mikaelian lens in w-space ($w=u + iv$) $n_w = 1/\cosh(mu)$ under the mapping $z = \exp(w)[x = \exp(u)\cos v, y = \exp(u)\sin v]$, the refractive index of generalized MFE lens in z space ($z=x + i\,y$) is

$$n_z = \frac{2r^{m-1}}{1+r^{2m}} \quad (r = \sqrt{x^2 + y^2}), \quad (1)$$

where *m* is an geometric integer [7, 15]. It becomes the MFE lens when *m*=1, and other index profiles of absolute instruments are available for other integers m, i.e. the generalized MFE lens [19]. In Fig. 1(a) the refractive index distribution and rays of the Mikaelian lens in *w* space for *m*=1 in Eq. (1) are shown, calculated by a finite element simulation software. In w space, the horizontal axis is *u* axis, and the vertical axis is *v* axis. The refractive index of Mikaelian lens decreases monotonically with increasing radius | *u* |, and the minimum value reaches up to zero. The maximum is reached along v axis, *u*=0, to the value of unity. In panel (a) three light rays emit in different incidence angles from the source point (2, -8) and advance periodically until converge to discrete foci located at *u*=±2. Whereas panel (b) shows the self-imaging effect of the MFE lens under the conformal mapping above, with a refractive index distribution of $n_z = 2/(1 + r^2)$ where the complex plane is denoted as $z=x+i\,y$. Note that we will truncate the infinite MFE into a finite one in our numeric investigation. The refractive index of the central point is 2, and gradually goes to zero when *r* increases. Within *z* plane the light rays propagate along the circles. In panel (b), the source point is (2, 0), and the image point is $(-1/2, 0)$, which aligns with the self-image effect in a graded index medium.

(a)

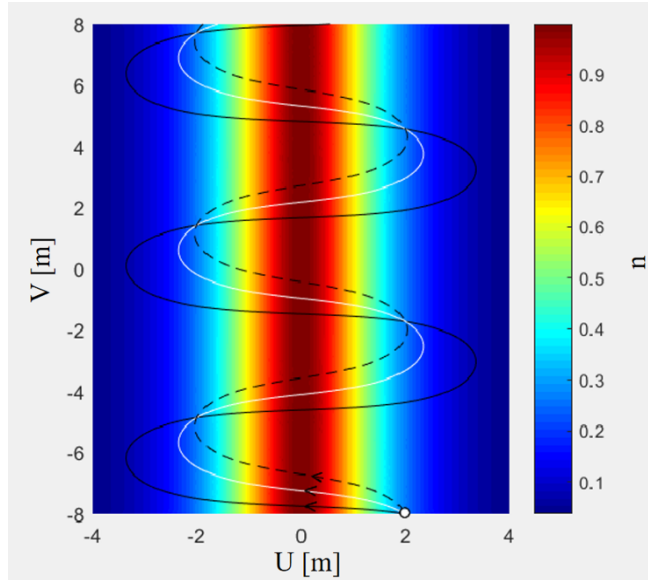

(b)

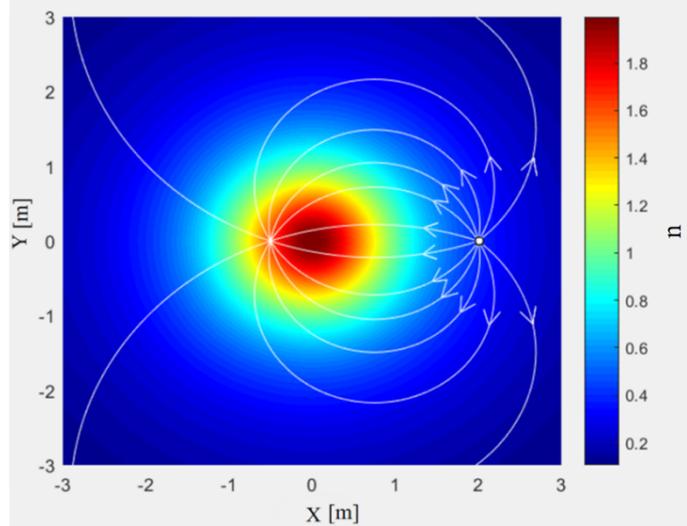

Fig. 1: Refractive index distribution of (a) Mikaelian lens and (b) MFE lens and their ray trajectories. The background color plot indicates the refractive index distribution. (a) The black, white and dashed lines in the figure denote the ray trajectories at different angles, forming 15°, 45°and 75°with u-axis. The source is at (2, -8) denoted by white dot. (b) The white curves denote the ray trajectories originating from (2,0).

Moreover, it takes more than a self-focusing lens such as MFE to achieve super-resolution imaging. Self-focusing lens such as MFE enables replicating from sources to images from either perspective of wave optics or geometrical optics. However, it remains open how one achieves subwavelength imaging in such an absolute instrument and many arguments around it were built such as [13, 20-25]. One of the main issues lies on the role of wave drain/sink and how it contributes to subwavelength imaging. Furthermore, a recent experiment in liquid setup provides further insights on super-resolution to which evanescent waves other than wave drains may contribute [26]. Before coming to the numerical results, we remark on the super-resolution and wave drain to clarify our simulation setup. The resolution of the imaging is in principle bound by diffraction limit, which obscures fine imaging details. Since diffraction limit is kept without drains in a MFE device [27], to achieve subwavelength resolution drains are required to embed within the MFE lens as outlets to absorb energy. If a drain can fully absorb the incident waves without generating reflecting or scattering waves, its radius should remain in the subwavelength scale [17, 14]. In principle, the resolution could be improved infinitely with the perfect drains fine enough in size, but is limited in realization by the finite size of the drain.

To investigate the drain effect, we set up a port form of wave drains in simulation to emulate the drain effects in a consistent and pragmatic manner, which also aligns with the known simple drain theory and its understanding on the drain's role in imaging [17, 14]. In simulation we shall pinpoint the imaging resolution in scale of the operating wavelength in the case of single-frequency wave solution. Also a mirror is added as the boundary of MFE $r$=1 where the refractive index reaches unity with a radius of 1m to fold the truncated region all mapped inside under affine mapping [5, 27], shown in Fig.

2(a). The light trajectories in panel (a) are then bound within the disk due to reflection on the boundary, and re-focus onto the image points, shown by the schematic in panel (b). For tunable ports as sources and drains show in Fig. 2(b), the radius of the source, port 1 and the drain, port 2 in Fig. 2(b) is chosen as 0.01m, and the perfect-electric-conductor (PEC)-backed port could either outflow waves from or inflow waves [17] with switchable wave excitation, and the wavenumber set for air in our simulation model, and we calculate only the electric field polarized along z direction. Our schematic uses two-dimensional MFE not only for its conceptual simplicity but also for its practical benefits to evanescently connect source and drain which will not disturb the exact profile of MFE. The energy flows throughout port 1 so that the electric field amplitude inside the ports suddenly drops to zero.

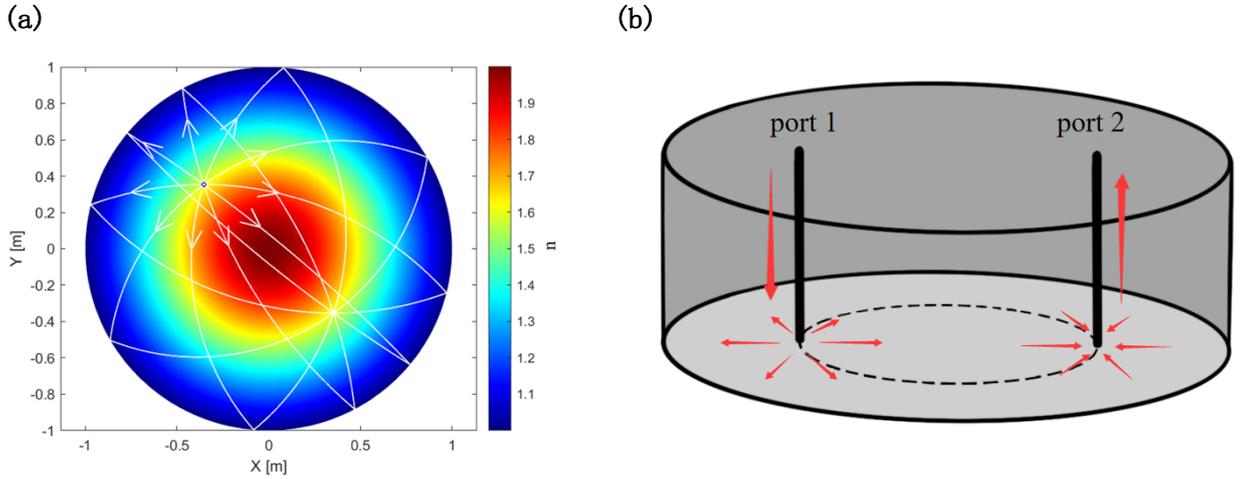

Fig 2: (a) Refractive index distribution in colour for the mirrored MFE lens and the light trajectories in black curves. The source is at ($\sqrt{2}/2, -\sqrt{2}/2$), and the image point is at ($-\sqrt{2}/2, \sqrt{2}/2$). (b) Schematic diagram of the subwavelength imaging principle in a two-dimensional MFE lens, where the red arrows indicate the energy flow direction out of the source and into the drain, respectively.

### A. Single frequency simulation for a *drained* MFE lens

To understand the role of drains in subwavelength imaging, sets of numerical simulation are performed with varying positions and number of ports in Subsec. A. In Fig. 3(a) an amplitude peak occurs at the anticipated image point which is the antipodal of the source point, but with a diffraction-limited resolution. The FWHM near the image point in Fig. 3(a) is calibrated as $0.394\lambda$ in scale of the working wavelength ($\lambda$=3m), still bounded by diffraction limit. A drain, port 2 is also set at the image point in Fig. 3 (b), in which a sharp increment occurs near port 2, but notably inside both of the ports, the electric fields just vanish. We stress that for a drain the wave excitation is switched off and therefore such field enhancements near it are solely due to excitation from other fields and sources in simulation. The FWHM for the drain, port 2 is well reduced to subwavelength scale as 0.0091λ. This shows the tell-tale hallmark of *super-resolution* beyond diffraction limit due to the wave port.

(a)

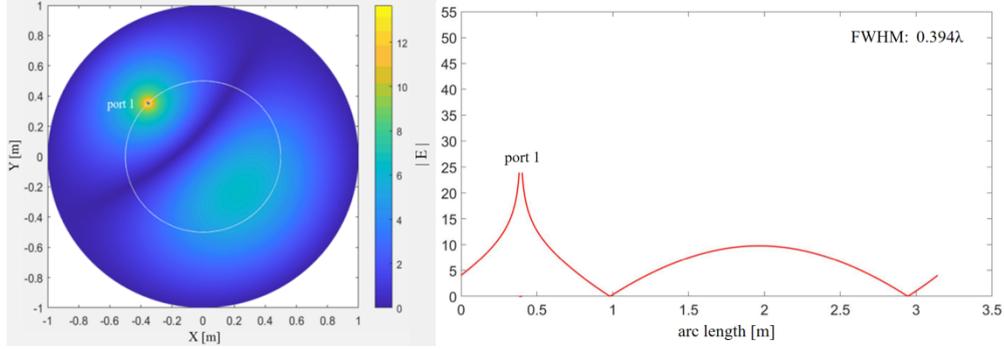

(b)

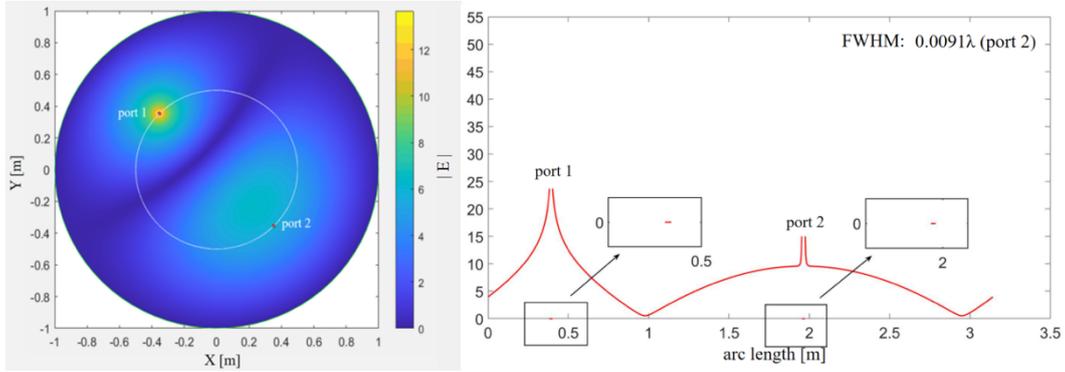

Fig. 3: Electric field amplitude of the MFE lens and on a circle of radius of 0.5 meters are shown for (a) without drain and (b) with a drain. The electric field source is realized by a wave port in simulation, and the field input sets as 1 A/m. (a) Without drain at the image point: the left panel shows the electric field distribution in the MFE lens, and the right figure plots the electric field distribution along the white circle in panel(a), in which the internal field of the ports both vanish to 0. (b) With a drain at the image point: fields at both the source port and the drain port are always 0.

In order to further investigate the roles of the wave ports for subwavelength imaging, two more ports are set up near port 1 and port 2, respectively, and the imaging fields in different cases are simulated, when we position the source and drain with different numbers and distances.

Firstly, one source and three drains are set up: two drains port 3 and port 4 are placed on both sides of port 2, with the distances from port 3 and port 4 to port 2 as 0.05 (m) and 0.39 (m), and all other settings unchanged, as shown in Fig. 4(a). For the closer drain, point 3 to the image point in figure also shows the resolution increment, but relatively weaker than the antipodal port 2, while the further drain port 4 far produces a slow wave trough limited by wave diffraction, probably due to the long distance from the image point. Only the FWHM of port 2 and port 3 are calculated to achieve subwavelength diffraction, which are 0.0062λ, 0.0065λ respectively. Second, three sources and one drain are set, and the distances between the two additional sources, port 5 and 6 with respect to port 1 are 0.05 m and 0.26 m respectively, shown in Fig. 4 (b). The optimal resolution occurs at port 2 is calculated as $0.0158\lambda$. Finally, three sources and three drains are set simultaneously in Fig. 4 (c). The electric field amplitude in panel (c) appears close to the superposition of those in panels (a) and (b), and only the peak field of the electric field changes, which does not change the general shape. The FWHM at port 2 and port 3 are calculated also as $0.0062\lambda$, $0.0065\lambda$, both of which goes beyond the diffraction limit.

To summarize Subsec. 3A, a wave drain contributes to subwavelength imaging at the location of antipodal point and multiple drains contribute to finer resolutions than a single drain. Moreover, increasing the number of source ports does not contribute to the super-resolution effect in our MFE lens. We also note that the vanished electric fields inside both source and drain ports only represent an idealised case which may complicate further in pragmatic experiments.

(a)
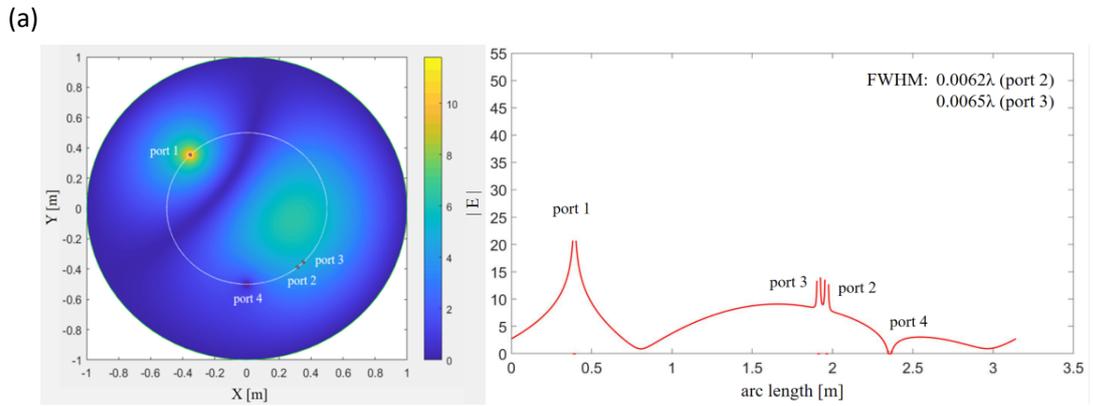

(b)
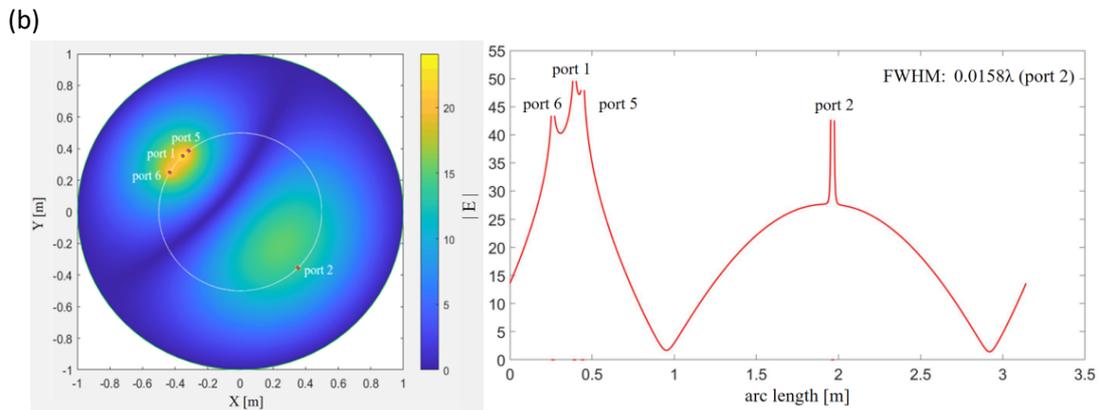

(c)

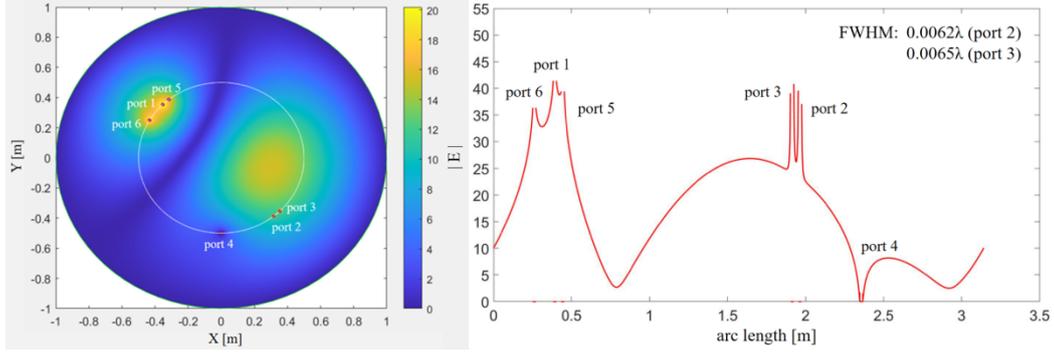

Fig. 4: Electric field amplitude when placing different numbers of source and drain ports (a-c). The left panels show the overall electric field distribution in the MFE lens, and the right ones give field profiles on the white curves in the left.

## B. Transmission spectra for a drained MFE lens

Instead of giving field distributions induced from a single-frequency wave source as done in Subsec. A, we now investigate the transmission spectrum for the source-to-drain energy flow in a mirrored MFE lens and find that the lens resolution varies with the frequency, in line with previous theory [14].

For frequency domain, the wavelength is chosen to vary from 0.5 to 5 meters, with the simple form of a single drain versus a single source, as shown in Fig. 5(a). We set up two cases to look at the energy transmission spectra when the drain is positioned in the aligned and the misaligned positions $\theta = 0$ and $0.1$, with θ representing the angle of the drain deviation from the oppositely-aligned position. For misaligned drain, the theoretical transmission spectra T is calculated by Eq. (2) for σ=4, where k stands for the wave number, and $P_\nu(\cos\theta)$ the Legendre function in order ν in Eq. (3).

$$T = \left| \frac{2i\sigma \sin\nu\pi\, P_\nu(\cos\theta)}{P_\nu(\cos\theta)^2 - (\cos\nu\pi - i\sigma \sin\nu\pi)^2} \right|^2, \quad (2)$$

With

$$\nu = \frac{1}{2}\left(\sqrt{4k^2 + 1} - 1\right). \quad (3)$$

For the perfect aligned drain, the transmission formula can be simplified to Eq. (4) below,

$$T = \frac{1}{\cos^2\nu\pi + a^{-2}\sin^2\nu\pi}, \quad (4)$$

With

$$a = \frac{8}{g^2 k + 16 g^{-2} k^{-1}}, \quad (5)$$

where g denotes the coupling coefficient of cable out of the MFE lens [14]. The transmission spectra of Eqs. (2-3) are calculated, and compared with the transmission curves simulated, in Fig. 5 (b-c) respectively. For the aligned drain $\theta=0$, which represents that the drain is in exact opposition to the source, the transmission reaches unity for an integer ν, the so-called notch frequency [10, 24]. Moreover, the simulation transmission also reaches the peaks for an integer ν, though up to less than 0.9, and the trough value is also smaller than theoretical formula [Eq. (4)]. We attribute the underplay to the transmission loss coupling wave ports in our simulation. When the drain is misaligned off the antipodal point to the source, e.g. $\theta=0.1$ rad, the theoretical transmission drops to Minano dip [11, 12] for an integer ν, with a symmetric peak on both sides of the notch frequency [10, 24]. Along the simulation curve, transmission also drops at the notch frequency, but the peaks on both sides of the notch frequency are less symmetric, with the peak on the right side of the notch larger than the left. In panel (f-g), we also present the field distributions for $\nu = 8.25$ with the drain of port 2 placed aligned and misaligned, respectively [the maximum transmission for the aligned drain occurs instead at $\nu = 8.03$ in panel (b)]. These two panels indicate that the misaligned drain (c) can capture less energy than the aligned one (b). To further understand the coupling mechanism between the two ports, reflection in port 1 and transmission in port 2 are also calculated compared with theoretical predictions in [14], given in panels (d-e) respectively. The curves in port 1 and 2 both indicate the coupling mechanism goes less as the misaligned angle $\theta$ increases. We are then assured that, our wave simulations using the simple ports as source and drain, successfully capture the coupling feature of Minano dip [11, 12] in scanning imaging of optical absolute instruments [17].

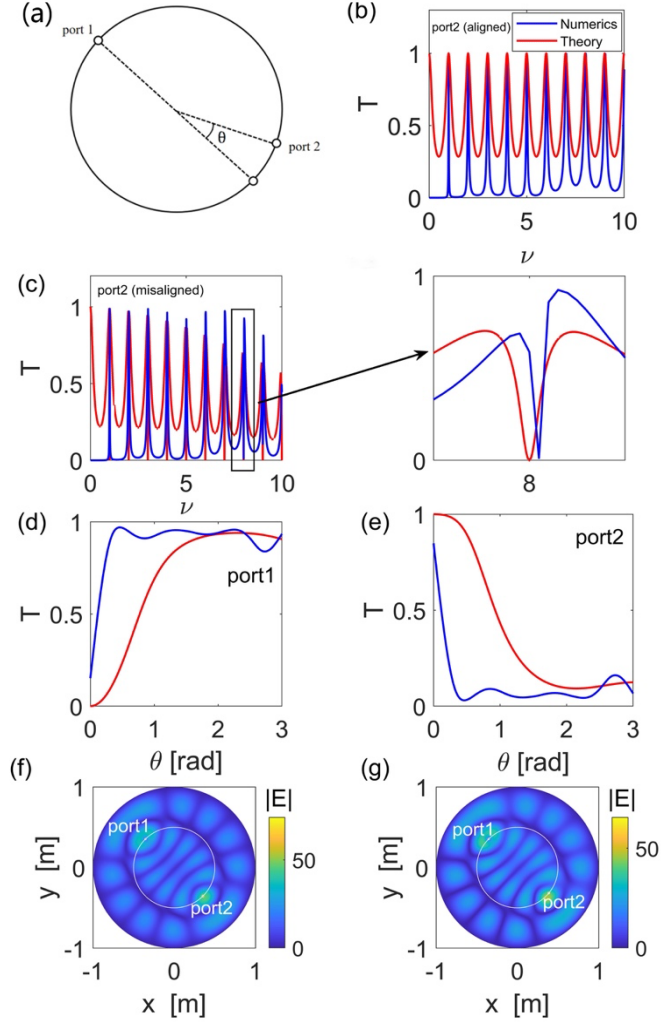

Fig. 5: One source, one drain. (a) A two-dimensional schematic diagram of the model, with a source as port 1 and a drain as port 2 separated by angle $\theta$ to the aligned antipodal position. (b-c) Transmission T with respect to frequency $\nu$ in (b) $\theta = 0$, (c) $\theta = 0.1$rad. (d-e) Reflection for port 1 (d) and Transmission for port 2 (e) with respect to angle $\theta$ to the aligned drain position for $\upsilon = 0.95$, $\sigma=1$. Red curves in panel (b-e) are theoretical curves: (b) for Eq. (3); (c) for Eq. (2). Blue curves in panel (b-e) for simulation. (f-g) Electric field distribution for $\nu = 8.25$ for port 2 placed in panel (b) and (c).

To further check the interaction between drains and sources, we also set up the cases of one source with two drains, for which the explicit derivation for transmission coefficients in Appendix A. We simulate the transmission spectra in Fig. 6 for two drains to demonstrate the interaction effect between them. The simulation spectra give similar trends compared with theoretical curves [14] but with damped amplitudes in both ports shown in panel (b) and (c) respectively. Moreover, the transmission spectra via the misaligned port 3 in (c) give null values at resonance as expected, but also manifest as Minano dips distinctly reminiscent of Fig. 5(c). This difference can be attributed to the less coupling efficiency of wave ports in our simulation compared with previous theoretical predication.

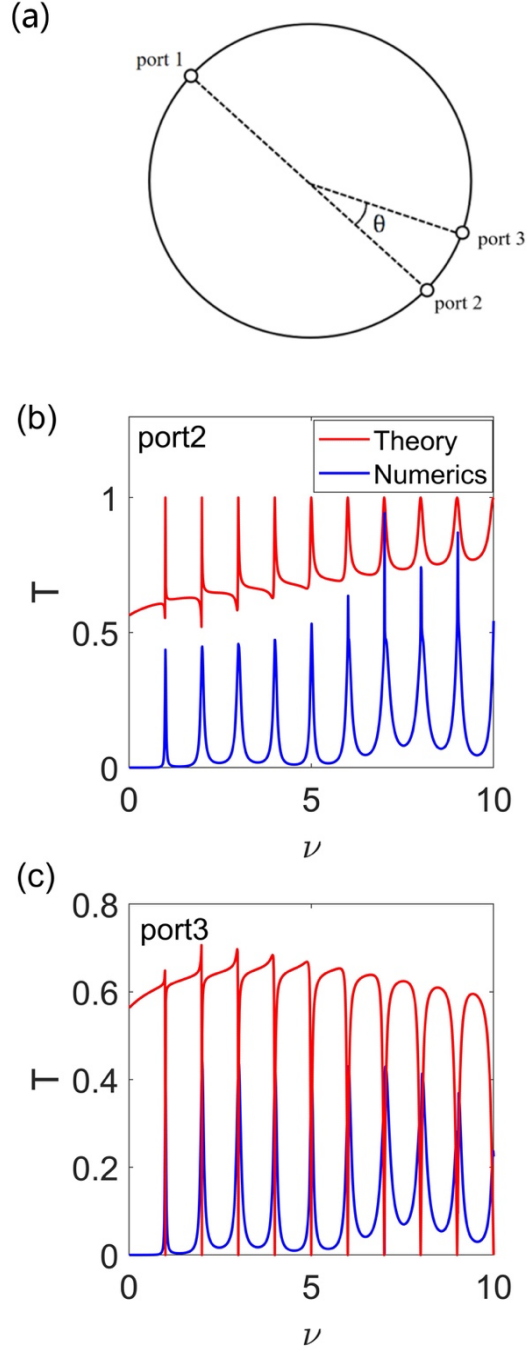

Fig. 6: One source, two drains. (a) A schematic with a source as port 1 and two drains as port 2 and 3 separated by angle $\theta = 0.1$rad. (b) Transmission of port 2. (c) Transmission of port 3. (b-c) Transmission T with respect to frequency $\nu$ in (b) port 2 with $\theta = 0$, (c) port 3 with $\theta = 0.1$rad. Red curves in panel (b-c) are the theoretical curves for Eqs. (A7-A8) respectively. Blue curves in panel (b-c) for simulation.

Furthermore we go for the case of two sources, two drains. Two wave sources of port 1 and 2, and two drains of port 3 and 4 are set antipodally as in Fig. 7(a), and the transmission spectra are calculated with respect to misaligned angle distance in Fig. panels (b-d), where theoretical spectra are also compared [also cf. Appendix A].

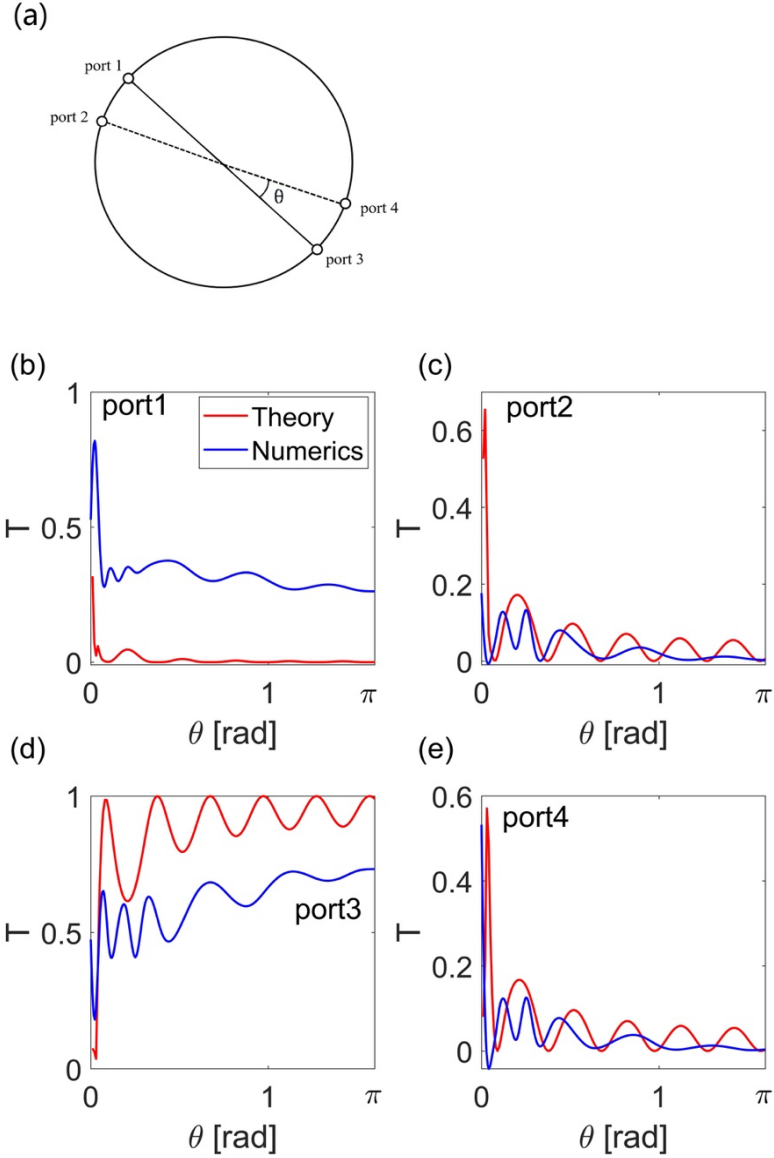

Fig. 7: Two sources, two drains. (a) A two-dimensional schematic diagram of the model, with two sources as port 1 and 2 and two drains as port 3 and 4. Only port 1 is excited, and two sources are separated by the angle θ to two aligned drains for ν = 9.984, σ=1. (b-e) Transmission T with respect to angle distance $\theta$: (b) Reflection in port 1, (c) Reflection of port 2. (d) Transmission of port 3. (d) Transmission of port 4. Red curves in panel (b-e) are the theoretical curves from Eq. (A11). Blue curves in panel (b-e) for simulation.

We observe that for the aligned drain, port 3 in panel (d) fires only when such a distance becomes farther away than less a half wavelength though it transmits less than theory, while the misaligned port 4 in panel (e) actually transmits closely to port 2 in panel (c), as expected. This observation concurs that subwavelength distance is irresolvable as Ref. [14] finds. And the fired port 1 in panel (b) even reflects more substantially in our simulation than in theory. This feature may also relate to the port coupling mechanism from wave fields.

## 3. Conclusion

In this paper, we numerically adopt a simple port model to realize both source and drain in a two-dimensional version of optical absolute instrument: a mirrored MFE lens. Our wave simulation assures that the coupling nature for wave source and drain applies correctly in the picture of scanning imaging for absolute instrument. Further details of wave drains remain worth looking at as well as their interaction and coupling with the wave fields outside. Our work is hoped to contribute to the still ongoing experimental efforts to realise geodesic MFE waveguides [28] and GRIN effective media [29, 30, 31] inspired from absolute instruments. We look forward to more surprising results of scanning near fields from far-field distances under the scene of simple-yet-universal wave physics [32].

## Reference (longlist)

Author contribution:

Q. S., C. G. worked on the simulation problem, processed the data, plotted all figures, and wrote the initial manuscript. M. L. and C. G. contributed to coding, data post-processing and plotting figures. X. Z. contributed to project management. Y. L. proposed the idea, interpret the results, supervised the project, and rewrote the whole manuscript. L. J., M. W., and Z. W. co-supervised the project progress and contributed to the result interpretation. L. J. also revised the paper, and H. X. contributed to Fig. 5(f-g).

Acknowledgments:

Y. L. thanks Hubei Key Laboratory of Ferroelectric and Piezoelectric Materials and Devices for its fruitful platform. We are supported by Young Scientist Fund [NSFC11804087] and Natural National Science Foundation [NSFC12047501]; Science and Technology Department of Hubei Province [2022CFB553, 2022CFA012, 2018CFB148]; Educational Commission of Hubei Province of China [T2020001]; Hubei University [X202210512039, 030-017643].

**Disclosure.** The authors declare no conflicts of interest.

**Data availability.** Data underlying the results presented in this paper are not publicly available at this time but may be obtained from the authors upon reasonable request.

**Appendix A.**

We set port 1 as the source, port 2 as the aligned drain and port 3 the misaligned one. According to Eq. (5) in [14], $z_1 = 0, z_2 = 1, z_3 = \cot \delta/2$, as in Fig. 6(a).

$$\zeta_{12} = \zeta_{21} = \zeta_1(z_2) = \frac{|z_2'|^2 - 1}{|z_2'|^2 + 1} = \frac{|z_2|^2 - 1}{|z_2|^2 + 1} = 1. \quad (A1)$$

$$\zeta_{13} = \zeta_{31} = \zeta_1(z_3) = \frac{|z_3'|^2 - 1}{|z_3'|^2 + 1} = \frac{|z_3|^2 - 1}{|z_3|^2 + 1} = \cos \delta. \quad (A2)$$

$$\zeta_{23} = \zeta_{32} = \zeta_2(z_3) = \frac{|z_3'|^2 - 1}{|z_3'|^2 + 1} = -\cos \delta. \quad (A3)$$

The coefficients for outgoing and incoming amplitudes of the three ports are related via scattering matrix **S**:

$$\begin{pmatrix} a_1' \\ a_2' \\ a_3' \end{pmatrix} = \mathbf{S} \begin{pmatrix} a_1 \\ a_2 \\ a_3 \end{pmatrix}, \quad (A4)$$

Where $\mathbf{S} = \mathbf{W^{-1}W^*}$, of which the asterisk sign symbols for complex conjugate and

$$\mathbf{W} = \begin{pmatrix} \cos\nu\pi - i\sigma\sin\nu\pi & P_\nu(\zeta_{12}) & P_\nu(\zeta_{13}) \\ P_\nu(\zeta_{21}) & \cos\nu\pi - \cos\nu\pi\, i\sigma\sin\nu\pi & P_\nu(\zeta_{23}) \\ P_\nu(\zeta_{31}) & P_\nu(\zeta_{32}) & \cos\nu\pi - i\sigma\sin\nu\pi \end{pmatrix}. \quad (A5)$$

Thus substituting $a_1 = 1, a_2 = a_3 = 0$, one obtains all three outgoing amplitudes $a_1', a_2', a_3'$ as below,

$$a_1' = -\frac{(\cos\nu\pi - i\sigma\sin\nu\pi)[(\sigma^2-1)\sin^2\nu\pi - (P_\nu(+\cos\delta))^2 - (P_\nu(-\cos\delta))^2] + 2P_\nu(+\cos\delta)P_\nu(-\cos\delta)}{(\cos\nu\pi - i\sigma\sin\nu\pi)[(\sigma^2+1)\sin^2\nu\pi + (P_\nu(+\cos\delta))^2 + (P_\nu(-\cos\delta))^2 + i\sigma\sin 2\nu\pi] - 2P_\nu(+\cos\delta)P_\nu(-\cos\delta)}, \quad (A6)$$

$$a_2' = \frac{4i\sigma\sin\nu\pi \cdot [\cos\nu\pi - i\sigma\sin\nu\pi - P_\nu(+\cos\delta)P_\nu(-\cos\delta)]}{(\cos\nu\pi - i\sigma\sin\nu\pi)[2(\sigma^2+1)\sin^2\nu\pi + 2(P_\nu(+\cos\delta))^2 + 2(P_\nu(-\cos\delta))^2 + 2i\sigma\sin 2\nu\pi] - 4P_\nu(+\cos\delta)P_\nu(-\cos\delta)}, \quad (A7)$$

$$a_3' = \frac{4i\sigma\sin\nu\pi \cdot [P_\nu(+\cos\delta)(\cos\nu\pi - i\sigma\sin\nu\pi) - P_\nu(-\cos\delta)]}{(\cos\nu\pi - i\sigma\sin\nu\pi)[2(\sigma^2+1)\sin^2\nu\pi + 2(P_\nu(+\cos\delta))^2 + 2(P_\nu(-\cos\delta))^2 + 2i\sigma\sin 2\nu\pi] - 4P_\nu(+\cos\delta)P_\nu(-\cos\delta)}. \quad (A8)$$

For two source and two drain, as [14] we put $\theta_1 = \pi, \theta_2 = \pi + \delta, \theta_3 = 0, \theta_4 = \delta, \phi_m = 0$, so the port coordinates in complex plane are

$$z_1 = \cot\frac{\pi}{2} = 0, z_2 = -\tan\frac{\delta}{2}, z_3 = \infty, z_4 = \cot\frac{\delta}{2}. \quad (A9)$$

We then work out for explicit elements of scattering matrix,

$$\zeta_{12} = \zeta_{34} = -\cos\delta, \zeta_{13} = \zeta_{34} = 1, \zeta_{14} = \zeta_{23} = \cos\delta, \quad (A10)$$

which just fulfills the geometrical symmetry by inspection similar to Eq. (A5). So four outgoing amplitudes can be obtained explicitly for each of the four ports again similar to Eq. (A4):

$$\begin{pmatrix} a_1' \\ a_2' \\ a_3' \\ a_4' \end{pmatrix} = \mathbf{S} \begin{pmatrix} a_1 \\ a_2 \\ a_3 \\ a_4 \end{pmatrix}, \quad (A11)$$

which are too lengthy to type here in text, but are readily available via using a computer algebra software taking care of matrix operation such as Mathematica.